\documentclass[11pt]{article}

\usepackage{amsmath}
\usepackage{graphicx}
\usepackage{amsfonts}
\usepackage{amssymb}
\usepackage{epsfig}
\usepackage{color}
\usepackage{psfrag}
\usepackage{epstopdf}

\newcommand{\EQ}{\begin{equation}}
\newcommand{\EN}{\end{equation}}
\newcommand{\be}{\begin{equation}}
\newcommand{\ee}{\end{equation}}
\newcommand{\bea}{\begin{eqnarray}}
\newcommand{\eea}{\end{eqnarray}}

\DeclareMathOperator{\sgn}{sgn}

\begin{document} \setcounter{page}{0}
\topmargin 0pt
\oddsidemargin 5mm
\renewcommand{\thefootnote}{\arabic{footnote}}
\newpage
\setcounter{page}{0}
\topmargin 0pt
\oddsidemargin 5mm
\renewcommand{\thefootnote}{\arabic{footnote}}
\newpage
\begin{titlepage}
\begin{flushright}
\end{flushright}
\vspace{0.5cm}
\begin{center}
{\large {\bf Critical lines in the pure and disordered $O(N)$ model}}\\
\vspace{1.8cm}
{\large Gesualdo Delfino and Noel Lamsen}\\
\vspace{0.5cm}
{\em SISSA and INFN -- Via Bonomea 265, 34136 Trieste, Italy}\\
\end{center}
\vspace{1.2cm}

\renewcommand{\thefootnote}{\arabic{footnote}}
\setcounter{footnote}{0}

\begin{abstract}
\noindent
We consider replicated $O(N)$ symmetry in two dimensions within the exact framework of scale invariant scattering theory and determine the lines of renormalization group fixed points in the limit of zero replicas corresponding to quenched disorder. A global pattern emerges in which the different critical lines are located within the same parameter space. Within the subspace corresponding to the pure case (no disorder) we show how the critical lines for non-intersecting loops ($-2\leq N\leq 2$) are connected to the zero temperature critical line ($N>2$) via the BKT line at $N=2$. Disorder introduces two more parameters, one of which vanishes in the pure limit and is maximal for the solutions corresponding to Nishimori-like and zero temperature critical lines. Emergent superuniversality (i.e. $N$-independence) of some critical exponents in the disordered case and disorder driven renormalization group flows are also discussed.
\end{abstract}
\end{titlepage}

\newpage
\tableofcontents

\section{Introduction}
Critical behavior in statistical systems is controlled by fixed points of the renormalization group. These are points in coupling space where the divergence of the correlation length allows for scale invariance to emerge. Fixed points are naturally described within the field theoretical framework, which in particular allows to see how their invariance properties actually extend, beyond scale transformations, to conformal transformations (see e.g. \cite{Cardy_book}). The charm and, at the same time, the difficulty of critical phenomena is that in general they correspond to non-trivial fixed points of the renormalization group, i.e. to interacting field theories that can only be studied within some approximation. An important exception is represented by the two-dimensional case, where conformal symmetry becomes infinite dimensional and conformal field theory (CFT) has provided a plethora of exact results extending to multi-point correlation functions at non-trivial fixed points \cite{BPZ,DfMS}. While this is a fantastic achievement, there are cases in two dimensions in which the sophisticated tools of CFT (e.g. differential equations for correlation functions) are not available, or have been so far too difficult to implement. In this respect, the idea of implementing the infinite dimensional conformal symmetry in a basis of particles rather than fields \cite{paraf,fpu} has proved quite useful. The resulting scale invariant scattering approach relies on few basic ingredients, i.e. elasticity of scattering processes, crossing symmetry and unitarity of the scattering matrix. This minimality of means makes the formalism very general and yields exact equations that allow to explore the space of fixed points with a given internal symmetry. 

For the permutational symmetry $S_q$, characteristic of the $q$-state Potts model, this exploration was performed in \cite{DT1} and revealed new results. In particular, the maximal value of $q$ allowing for a fixed point was found to be\footnote{The fixed points obtained in \cite{DT1} contain fields with real scaling dimensions and are not related to the complex CFT's with $q>4$ recently studied in \cite{GRZ}.} $5.56..$, instead of the previously expected value $4$. This leaves room for a second order phase transition in a $q=5$ antiferromagnet for which lattice candidates exist \cite{DHJSS,Huang}. A line of fixed points with $q=3$ and central charge $1$ was also predicted for which a lattice realization has recently been found \cite{LDJS}. 

A further, particularly remarkable feature of the scale invariant scattering formalism in two dimensions is that it extends to the problem of quenched disorder \cite{random}, i.e. to those ``random'' fixed points that had seemed out of reach for exact methods. In particular, for the random bond Potts model it was then possible to see analytically for the first time the softening of the phase transition by disorder above $q=4$ \cite{random}, a phenomenon expected on rigorous grounds \cite{AW} (see also \cite{HB}). Quite unexpected was instead the emergence of a mechanism allowing for the $q$-independence of the correlation length critical exponent $\nu$, with the magnetization exponent $\beta$ remaining $q$-dependent \cite{random}. This phenomenon of partial superuniversality finally accounts for the evidence emerged from a variety of investigations \cite{CFL,DW,KSSD,CJ,CB2,OY,JP,Jacobsen_multiscaling,AdAI} that $\nu$ does not show any appreciable deviation from the value 1 up to $q=\infty$.

Recently \cite{DL1} we have shown how the scale invariant scattering approach can be applied to the $O(N)$ vector model with quenched disorder and gives global and exact access to a pattern of fixed points that previously had been the object of perturbative (for weak disorder) \cite{Shimada} and numerical \cite{SJK} studies. In this paper we give all the solutions of the fixed point equations (table~\ref{solutions} below) and discuss their physical properties. In particular, we show how the critical lines parametrized by $N$ are located with respect to each other in the space of parameters. This is interesting already for the case without disorder (``pure''), for which we exhibit how the passage from the critical lines for non-intersecting loops ($-2<N<2$) to the zero temperature critical points ($N>2$) is ``mediated'' by the Berezinskii-Kosterlitz-Thouless (BKT) line at $N=2$ (figure~\ref{pure_space}). Disorder introduces two more dimensions in parameter space (figure~\ref{total}), and the behavior of one of these disorder parameters, which we call $\rho_4$, as a function of $N$ allows us to divide the critical lines in three classes. The class with $\rho_4=0$ corresponds to the pure case, while a class with $\rho_4=1$ includes Nishimori-like multicritical points. A third class, with $\rho_4$ varying between 0 and 1, describes fixed points possessing a weak disorder limit. Renormalization group flows between different critical lines can be inferred and associated to the phase transition boundary in the disordered spin model (figure~\ref{flows}). 

The paper is organized as follows. In the next section we recall basic definitions of the lattice $O(N)$ vector spin model and then give the scale invariant scattering description of $O(N)$ symmetry in the replicated setting required for the study of disorder. In section~3 we give the solutions of the fixed point equations obtained from the scattering formalism, presenting first the pure case, then the replicated case, and finally the disorder limit of zero replicas. The physical properties of the solutions are then discussed in section~4 for the pure case and in section~5 for the disordered case. The last section contains a summary of results and some additional remarks.

\section{$O(N)$ model and scale invariant scattering}
\begin{figure}
\begin{center}
\includegraphics[width=5cm]{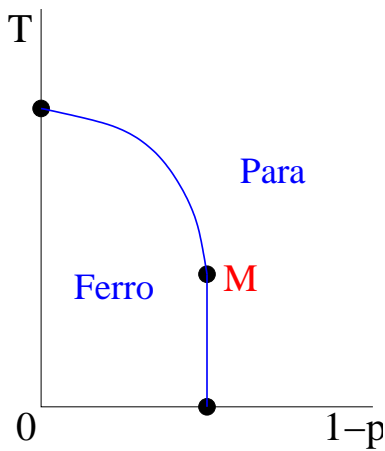}
\caption{Qualitative phase diagram and expected fixed points for the two-dimensional Ising model with the disorder distribution (\ref{pmJ}). $1-p$ is the amount of disorder with respect to the pure ferromagnet, and $M$ indicates the multicritical (Nishimori) point. 
}
\label{phd}
\end{center} 
\end{figure}

The $O(N)$ symmetry we study in this paper can be realized on a lattice considering the spin model defined by the Hamiltonian
\EQ
{\cal H}=-\sum_{\langle i,j\rangle}J_{ij}\,{\bf s}_i\cdot{\bf s}_j\,,
\label{lattice}
\EN
where ${\bf s}_i$ is a $N$-component unit vector located at site $i$, the sum is taken over nearest neighboring sites, and $J_{ij}$ are bond couplings. The pure case is that in which $J_{ij}=J$; $J>0$ for a ferromagnet. The disordered, or random, case (see e.g. \cite{Cardy_book}) corresponds instead to bond couplings drawn from a probability distribution $P(J_{ij})$, with the average over disorder taken on the free energy,
\EQ
\overline{F}=\sum_{\{J_{ij}\}}P(J_{ij})F(J_{ij})\,.
\EN
Most theoretical treatments of disorder rely on the fact that $F$ is related to the partition function $Z=\sum_{\{s_i\}}e^{-{\cal H}/T}$ as  $F=-\ln Z$, so that the identity
\EQ
\overline{F}=-\overline{\ln Z}=-\lim_{m\to 0}\frac{\overline{Z^m}-1}{m}
\EN
transforms the problem into that of $m\to 0$ replicas coupled by the disorder average. A particularly interesting disorder distribution is that mixing ferromagnetic and antiferromagnetic bonds in the form
\EQ
P(J_{ij})=p\delta(J_{ij}-1)+(1-p)\delta(J_{ij}+1)\,;
\label{pmJ}
\EN
the pure ferromagnet is recovered when the fraction $1-p$ of antiferromagnetic bonds goes to zero. Figure \ref{phd} qualitatively shows the phase diagram obtained by numerical simulations (see e.g. \cite{PHP,HPtPV}) for the two-dimensional Ising model with the disorder distribution (\ref{pmJ}). A ferromagnetic and a paramagnetic phase are separated by a phase boundary along which some fixed points of the renormalization group are located. Besides the pure fixed point and a zero temperature fixed point, the multicritical point known as Nishimori point \cite{Nishimori} is also present.

\begin{figure}
\begin{center}
\includegraphics[width=16cm]{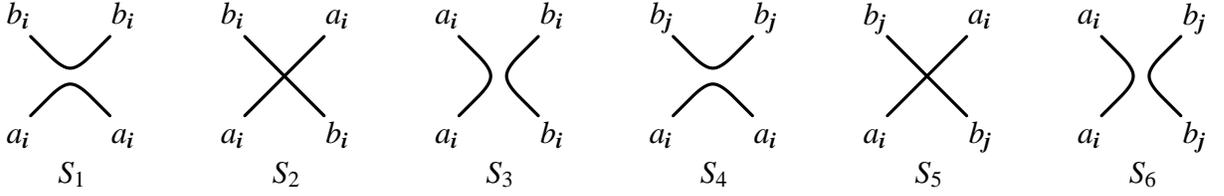}
\caption{Scattering amplitudes of the replicated $O(N)$ theory. Time runs upwards, indices $i$ and $j$ label different replicas. 
}
\label{ampl}
\end{center} 
\end{figure}

\begin{figure}
\begin{center}
\includegraphics[width=16cm]{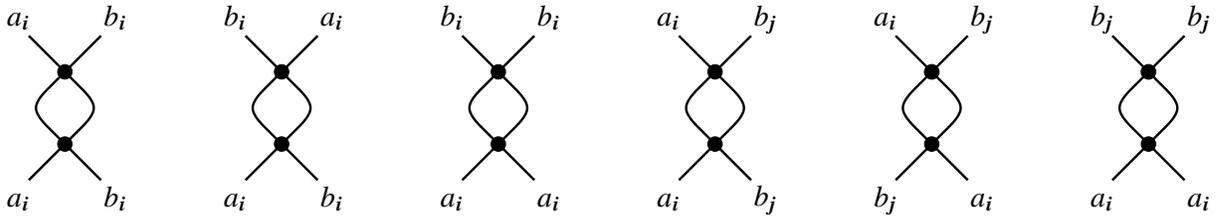}
\caption{Pictorial representations associated to the unitarity equations (\ref{u1})--(\ref{u6}), in the same order. The dotted scattering nodes stay for any of the amplitudes in figure~\ref{ampl}. The lower amplitude multiplies the complex conjugate of the upper amplitude, and summation is performed over all intermediate particle labels compatible with the assigned external labels.
}
\label{unitarity}
\end{center} 
\end{figure}

Our goal is that of identifying renormalization group fixed points for the case of $O(N)$ symmetry in two dimensions, both in the pure and in the disordered case. We exploit the fact that scale invariance implies an infinite correlation length, and allows the study of universal properties directly in the continuum, within a field theoretical framework. Since we consider systems that (after the disorder average, in the disordered case) are translationally and rotationally invariant, the corresponding field theory is the analytic continuation to imaginary time of a relativistically invariant quantum field theory. As shown in \cite{paraf} and \cite{random} for the pure and random case, respectively, these theories can be studied directly at fixed points exploiting the particle formalism. The first step is to identify the particle basis corresponding to the symmetry of the problem, and for $O(N)$ this is obtained through a vector multiplet representation of the particle excitations. Scale invariance in two (one space and one time) dimensions\footnote{See \cite{fpu} for an overview on fields, particles and critical phenomena in two dimensions.} implies that these particles are left- and right-movers with momentum and energy related as $p=\pm E$. In addition, within the replicated setting needed to deal with disorder, the excitations exist in each of the $m$ replicas. We will denote them as ${a_i}$, where $a=1,2,\ldots N$ labels the components of the vector multiplet, and $i=1,2,\ldots,m$ labels the replicas. In local field theories scale invariance gets enlarged to conformal invariance, and in two dimensions the latter is an infinite dimensional symmetry \cite{BPZ,DfMS}. Hence, there are infinitely many conserved quantities forcing the scattering of a right-mover with a left-mover to lead to final states still containing only a right-mover and a left-mover (``elasticity''). Moreover, scale invariance forces the scattering amplitudes to be energy independent. The product of two vectorial representations allows for the six scattering amplitudes shown in figure~\ref{ampl} \cite{DL1}. They amount to transmission and reflection of the particles in the same replica ($S_2$ and $S_3$, respectively), or in different replicas ($S_5$ and $S_6$); it is also possible that two identical particles annihilate and produce another pair in the same replica ($S_1$) or in a different replica ($S_4$). The amplitudes are related under exchange of space and time directions by crossing symmetry \cite{ELOP}, which yields the relations
\bea
S_1=S_3^{*} &\equiv &  \rho_{1}\,e^{i\phi}, \\
S_2 = S_2^* &\equiv & \rho_2,\\
S_4 = S_6^* &\equiv &  \rho_4\, e^{i\theta}, \\
S_5 = S_5^*&\equiv & \rho_5,
\eea 
and allows us to write the amplitudes in terms of the variables $\rho_1$ and $\rho_4$ non-negative, and $\rho_2$, $\rho_5$, $\phi$ and $\theta$ real. A final fundamental constraint for the amplitudes is provided by the requirement of unitarity of the scattering matrix, which is pictorially expressed in figure~\ref{unitarity} and produces the equations
\bea
&& \rho_1^2+\rho_2^2=1\,,  \label{u1}\\
&& \rho_1 \rho_2 \cos\phi=0\,,\label{u2}  \\
&& N \rho_1^2 + N(m-1)\rho_4^2 + 2\rho_1\rho_2 \cos\phi +2\rho_1^2\cos2\phi=0\,, \label{u3} \\
&& \rho_4^2 + \rho_5^{2}=1\,,\label{u4}\\
&& \rho_4 \rho_5 \cos\theta=0\,, \label{u5}\\
&& 2 N \rho_1 \rho_4 \cos(\phi-\theta) + N(m-2)\rho_4^2 + 2\rho_2\rho_4\cos\theta + 2\rho_1\rho_4\cos(\phi+\theta)=0\, \label{u6}. 
\eea
Equations (\ref{u1}) and (\ref{u4}) yield the restrictions
\bea
&& 0\leq\rho_1\leq 1\,,\hspace{1cm}-1\leq\rho_2\leq 1\,,
\label{c1}\\
&& 0\leq\rho_4\leq 1\,,\hspace{1cm}-1\leq\rho_5\leq 1\,,
\label{c2}
\eea
Notice that, while $N$ and $m$ are originally positive integers, they enter the equations as parameters that can be treated as taking real values. In particular, the disorder limit $m\to 0$ can be taken without difficulty.

It is not difficult to check that the superposition of two-particle states $\sum_{a,i}a_ia_i$ scatters into itself with the amplitude
\EQ
S=NS_1+S_2+S_3+(m-1)NS_4\,,
\label{singlet}
\EN
which has to be a phase by unitarity. Similarly, the superpositions $a_ib_i\pm b_ia_i$ and $a_ib_j\pm b_ja_i$ scatter into themselves with phases
\bea
\Sigma_\pm &=& S_2\pm S_3\,,
\label{Sigma}\\
\bar{\Sigma}_\pm &=& S_5\pm S_6\,,
\eea
respectively. 

It was shown in \cite{paraf,D_09} that the amplitude corresponding to the symmetry invariant scattering channel, which in the present case is the phase (\ref{singlet}), can be written as
\EQ
S=e^{-2i\pi\Delta_\eta}\,,
\label{phase}
\EN
where $\Delta_\eta$ is the conformal dimension of the chiral field $\eta$ that creates a right-moving particle. We recall that the conformal dimensions $(\Delta_\Phi,\bar{\Delta}_\Phi)$ of a field $\Phi(x)$ determine the scaling dimension $X_\Phi$ and the spin $s_\Phi$ as 
\bea
X_\Phi &=& \Delta_\Phi+\bar{\Delta}_\Phi\,,\\
s_\Phi &=& \Delta_\Phi-\bar{\Delta}_\Phi\,.
\eea
A field is chiral if one of its conformal dimensions vanishes. The energy density field $\varepsilon(x)$ and the spin field ${\bf s}(x)$, which are the other two main fields to be considered below, are instead spinless fields ($\Delta_\varepsilon=\bar{\Delta}_\varepsilon$, $\Delta_s=\bar{\Delta}_s$).

\section{Solutions of the fixed point equations}
The solutions of the equations (\ref{u1})--(\ref{u6}) correspond to renormalization group fixed points invariant under $O(N)$ transformations and replica permutations. In this section we give the solutions to these equations. We start with the pure case ($m=1$), and then move to the case of $m$ generic, which we finally specialize to the disorder limit $m=0$. The analysis will lead to the results for the pure and disordered solutions summarized in table~\ref{solutions}. In this table and in the whole discussion we take into account that equations (\ref{u1}) and (\ref{u4}) allow us to write
\bea
\rho_1 &=& \sqrt{1-\rho_2^2}\,,
\label{rho1}\\
\rho_5 &=& \pm\sqrt{1-\rho_4^2}\,,
\label{rho5}
\eea
so that the number of parameters characterizing the solutions is four. Two of them ($\phi$ and $\rho_2$) are sufficient to span the space of solutions for the pure case, while the remaining two ($\rho_4$ and $\theta$) are disorder parameters. The quadratic nature of the unitarity equations leads to solution doublings according to the sign of some parameters. While these doublings are not expected to be always physical, we will show in section~\ref{properties} that in some cases they are. 

\subsection{Pure case}
In the pure case only the amplitudes $S_1$, $S_2$, and $S_3$, which involve a single replica, are physical, and we have only equations (\ref{u1}), (\ref{u2}), and (\ref{u3}), the latter with $m=1$. Equivalently, the pure case can be regarded as that of $m$ decoupled replicas ($S_4=S_6=0$, $S_5=\pm 1$), and is obtained for $\rho_4=0$. Since the replicas are coupled by the disorder, it follows that $\rho_4$, to which we refer as ``disorder modulus'', gives a measure of disorder strength\footnote{For strong disorder, however, also the second disorder parameter $\theta$ becomes important, see below.}.

The solutions for the pure case are those of type $P$ in table~\ref{solutions}. The solutions $P1_\pm$ exist for any $N$ and correspond to non-interacting bosons/fermions. Indeed, scattering on the line mixes statistics and interaction, and the transmission amplitude $-1$ corresponds to free fermions. The solutions $P3_\pm$ exist only for $N=2$ and contain $\rho_1$ as a free parameter corresponding to the coordinate along a line of fixed points.

\subsection{Interacting replicas}
Interacting replicas require $\rho_4\neq 0$, and (\ref{u5}) shows that we can distinguish two classes of solutions -- one with $\cos\theta=0$ and one with $\rho_5=0$. Below we restrict our attention to solutions defined in intervals of $m$ containing $m=0$, which are those of interest for the purpose of taking the disorder limit. In writing the solutions we take into account that the equations fix the relative sign of $\sin\phi$ and $\sin\theta$, which we specify giving the value of 
\EQ
\gamma\equiv \sgn(\sin\phi \sin \theta)\,.
\label{gamma}
\EN

Up to sign doublings, there are two solutions in the class with $\cos\theta=0$. The first is defined for $N\in[-\sqrt{2}-1,-2]\cup[\sqrt{2}-1,\infty)$ and reads
\begin{equation}
\begin{split}
\rho_2 = 0,& \hspace{40pt} \rho_4 = |N-1|\sqrt{ \frac{N+2}{N\left( N m^2 + (N+1)^2(1-m) \right)} }, \\[1em]
 2 \cos\phi &= \pm \sqrt{\frac{4(1-m) - N(N-2)m^2}{(N+1)^2(1-m) + N m^2}}, 
\\[1em]
 - \sqrt{2} - 1 &\leq N \leq -2, \quad  -\tfrac{2}{|N|} \leq m \leq \tfrac{N^2 + 2N -1}{N}, \quad \gamma = +1 \\
 \sqrt{2} - 1 = N_{*} & \leq N < 1, \quad   m \geq \tfrac{N^2 + 2N -1}{N}, \quad \gamma = -1 \\
 1 &< N \leq 2, \quad  m \leq \tfrac{2}{N} , \quad \gamma = +1 \\
 N &> 2 , \quad \tfrac{2}{2-N} \leq  m \leq \tfrac{2}{N} , \quad \gamma = +1,
\end{split} 
\label{V1m}
\end{equation}
where, for different intervals of $N$, we specified the value of $\gamma$ and the allowed interval of $m$. The second solution with $\cos\theta=0$ is defined only for $N=0$ and has
\begin{align}
\label{V2m}
\rho_2 = \pm 1, \hspace{1cm}  0 \leq \rho_4 \leq 1\,;
\end{align}
it is $m$-independent and corresponds to a line of fixed points parametrized by $\rho_4$.

Up to sign doublings, the class with $\rho_5=0$ contains four solutions. The first is defined for any $N$ and reads
\begin{equation}
\begin{split}
\rho_1 &= \rho_4 = 1, \qquad 2 \cos\phi = \pm \sqrt{2 - N m}\,,  \\[0.5em]
2 \cos \theta &= \pm \dfrac{\sqrt{2 - N m} (N^2 - 1 + N(2-m) ) }{N^2 (1-m) +1}, \qquad -\tfrac{2}{|N|} \leq m \leq \tfrac{2}{|N|},
\end{split}
\label{S1m}
\end{equation}
with $\gamma = \pm 1$ when $(m - \overline{m}(N))\sgn(N)\gtrless 0$, where $\overline{m}(N) = 2 - N + \tfrac{1}{N}$ is the boundary along which $\gamma=0$. The second solution is also defined for any $N$, differs from the previous one only in $\theta$, and reads
\begin{equation}
\begin{split} 
\rho_1 &= \rho_4 = 1, \quad 2 \cos\phi = 2 \cos \theta  = \pm \sqrt{2 - N m}, \quad  \gamma = + 1, \quad -\tfrac{2}{|N|} \leq m \leq \tfrac{2}{|N|}.
\end{split}
\label{S2m}
\end{equation} 
The third and fourth solutions with $\rho_5=0$ are defined only for $N\leq 0$ and again differ from each other only for the value of $\theta$. We write them as 
\begin{equation}
\rho_1 = \sqrt{\frac{N(m-1)}{2-N}}, \quad \rho_2 = \pm \sqrt{\frac{2 - N m}{2 - N}}, \quad \cos \phi = 0, \quad \rho_4 = 1,
\label{S34m}
\end{equation}
\begin{equation}
\begin{split}
\cos \theta = \begin{cases}
\mp \sqrt{\frac{2- N m}{2 - N}} \frac{(N-1)\sqrt{N (m-1)(2 + N m)}+ N(2-m)}{2(N^2 (1-m) + 1)},  \\[0.9em]
\pm \sqrt{\frac{2- N m}{2 - N}} \frac{(N-1)\sqrt{N (m-1)(2 + N m)}- N(2-m)}{2(N^2 (1-m) + 1)}; 
\end{cases} 
\end{split}
\nonumber
\end{equation}
the upper solution for $\cos\theta$ has $\gamma=1$ and $-\frac{2}{|N|}\leq m\leq \frac{2}{2-N}$, while the lower solution has $\gamma = -\sgn\left ( m - \widetilde{m}(N) \right )$ and $-\frac{2}{|N|} \leq m \leq \min\left [\frac{2}{|N|}, 1 \right ]$, where $\widetilde{m}(N) = 2[(N-1)^2 - \sqrt{2N^2 -6N +5}]/(N^2 -2N)$.

\subsection{Disorder limit}
\label{m0}

We now take the limit $m\to 0$ of the results of the previous subsection and obtain the solutions of type $V$ and $S$ in table~\ref{solutions}. 

The solution ${V}1_{\pm}$ is the limit of (\ref{V1m}), and has $\gamma= -1$ for  $N_*  \leq N < 1$ and $\gamma=1$ otherwise. ${V}2_{\pm}$ coincides with (\ref{V2m}), which was $m$-independent. The solutions ${S}1_{\pm}$ and ${S}2_{\pm}$ are obtained from (\ref{S1m}) and (\ref{S2m}), respectively; $\gamma$ is negative only for $S1_\pm$ in the interval $1 - \sqrt{2} < N < 1 + \sqrt{2}$. Finally, the solutions ${S}3_{\pm}$  and ${S}4_{\pm}$ are obtained from (\ref{S34m}) and are specified in table~\ref{solutions} with
\EQ
f_\pm(N)=\dfrac{\sqrt{-N}(N\pm \sqrt{-2N} -1)}{\sqrt{2-N}(N^2 +  1)}\,;
\label{fN}
\EN
$\gamma$ is negative only for $S4_\pm$ in the interval $\widetilde{N} < N \leq 0$, where $\widetilde{N} \simeq - 0.839$ is the real root of the polynomial $N^3 -2 N^2 + 2$.

\begin{table}
\begin{center}
\begin{tabular}{c|c|c|c|c|c}
\hline 
Solution & $N$ & $\rho_2$ & $\cos\phi$ & $\rho_4$ & $\cos\theta$ \\ 
\hline \hline
$\text{P}1_{\pm}$ & $\mathbb{R}$ & $\pm 1$ & - & - & - \\ 
$\text{P}2_{\pm}$ & $[-2, 2]$ & $0$ & $\pm\frac{1}{2}\sqrt{2-N}$ & - & - \\ 
$\text{P}3_{\pm}$ & $2$ & $\pm \sqrt{1- \rho_1^2}$ & $0$ & - & -  \\[0.7em] 
\hline 
$\text{V}1_{\pm}$ & \parbox[m]{70pt}{\begin{equation*}\begin{split}[-1-\sqrt{2}, -2] \\ \cup \; [\sqrt{2}-1,\infty)\end{split}\end{equation*}} & $0$ & $\displaystyle \pm \frac{1}{|N+1|} $ & $ \displaystyle \left | \frac{N-1}{N+1} \right |\sqrt{\frac{N+2}{N}} $ & 0 \\ 
$\text{V}2_{\pm}$ & $0$ & $\pm 1$ & - & $[0, 1]$ & $0$ \\[0.7em] 
\hline 
$\text{S}1_{\pm}$ & $\mathbb{R}$ & $0$ & $\pm\dfrac{1}{\sqrt{2}}$ & $1$ & $ \pm \dfrac{N^2 + 2N -1}{\sqrt{2}(N^2 + 1)}$ \\[0.7em] 
$\text{S}2_{\pm}$ & $\mathbb{R}$ & $0$ & $\pm \dfrac{1}{\sqrt{2}}$ & $1$ & $\pm \dfrac{1}{\sqrt{2}}$ \\[0.7em] 
$\text{S}3_{\pm}$ & $(-\infty , 0]$ & $\pm \sqrt{\dfrac{2}{2-N}}$ & $0$ & $1$ & $\mp f_+(N)$  \\[0.7em]
$\text{S}4_{\pm}$ & $(-\infty , 0]$ & $\pm \sqrt{\dfrac{2}{2-N}}$ & $0$ & $1$ & $\pm f_-(N)$  \\
[0.7em] 
\hline 
\end{tabular} 
\caption{Solutions corresponding to renormalization group fixed points with $O(N)$ symmetry in the pure ($P$ type) and disordered ($V$ and $S$ types) cases. The functions $f_\pm(N)$ are given by (\ref{fN}).
}
\label{solutions}
\end{center}
\end{table}

\section{Properties of solutions for the pure case}
\label{properties}
\subsection{Critical lines of non-intersecting loops}
The solutions $P2_\pm$ have already been identified in \cite{paraf,DT1}; here we summarize the main steps of the derivation and the results. For $N=2$ these solutions coincide with the point $S_2=0$ of the solution $P3$ that, as we will see below, corresponds to a CFT with central charge $c=1$. Since the central charge grows with $N$, the CFT's describing the solutions $P2_\pm$ will have $c\leq 1$. In this subspace of CFT a main physical role is played by the ``degenerate'' primary fields $\Phi_{m,n}$ \cite{BPZ,DfMS} with conformal dimensions
\EQ
\Delta_{m,n}=\frac{[(p+1)m-pn]^2-1}{4p(p+1)}\,,
\label{deltamunu}
\EN
where $m,n=1,2,\ldots$, and $p$ determines the central charge through the relation
\EQ
c=1-\frac{6}{p(p+1)}\,.
\label{cc}
\EN
The energy density field $\varepsilon(x)$ is expected to be a degenerate field, and at $N=1$ for one of the two solutions should have the conformal dimension $\Delta_\varepsilon=1/2$ of the Ising model ($p=3$). This leads to the identification $\varepsilon=\Phi_{1,3}$, while the alternative choice $\Phi_{2,1}$ is found to correspond to the $q$-state Potts model. The requirement that the chiral field $\eta$ entering (\ref{phase}) is local\footnote{Two fields $\Phi_1$ and $\Phi_2$ are mutually local if the correlators $\langle\cdots\Phi_1(x_1)\Phi_2(x_2)\cdots\rangle$ are single valued.} with respect to $\varepsilon$ then leads to the identification $\eta=\Phi_{2,1}$, i.e. to the determination of $\Delta_\eta$ as a function of $p$. On the other hand, $\Delta_\eta$ is given as function of $N$ by the relation (\ref{phase}), in which $S=NS_1+S_3=-e^{3i\phi}$ for the solutions $P2_\pm$. For $N=1$ the solution we are discussing corresponds to Ising, i.e. to a free fermion theory with $S=-1$, and this selects $P2_-$. Comparing the two results for $\Delta_\eta$ (one as a function of $p$ and one as a function of $N$) we obtain the relation $N=2\cos\frac{\pi}{p}$ for the solution $P2_-$. A slightly more general analysis involving nondegenerate fields \cite{paraf} yields also $\Delta_s=\Delta_{1/2,0}$ for the conformal dimension of the spin field.

The simplest way to identify the solution $P2_+$ is to recall that the perturbation by the field $\Phi_{1,3}$ of the CFT's with central charge (\ref{cc}) yields (for one sign of the coupling) flows to infrared fixed points with central charge corresponding to $p-1$ \cite{Zamo_cth}. Since the $\Phi_{1,3}=\varepsilon$ perturbation preserves $O(N)$ symmetry, the infrared line of fixed points corresponds to $P2_+$, and has $N=2\cos\frac{\pi}{p+1}$. Together with (\ref{phase}), this relation yields $\Delta_\eta=\Delta_{1,2}$, a result differing from that for $P2_-$ for the interchange of the indices $m,n$. This interchange is preserved by the mutual locality arguments (which exploit the operator product expansion), and leads to $\Delta_\varepsilon=\Delta_{3,1}$ and $\Delta_s=\Delta_{0,1/2}$ for the critical line $P2_+$. The results for the solutions $P2_\pm$ are summarized in table~\ref{table_pure}. 

The solutions $P2_\pm$ are characterized by $S_2=0$, i.e. by the absence of intersection of particle trajectories (see figure~\ref{pure_space}). It must be recalled that it is possible to map (see e.g. \cite{Cardy_book}) the partition function of a $O(N)$ invariant ferromagnet on that of a loop gas,
\EQ
Z_\textrm{loops}=\sum_G K^{n_b}N^{n_l}\,,
\label{loops}
\EN
where the sum is over loop configurations $G$, $K$ is the coupling in the spin formulation, $n_b$ is the number of lattice edges occupied by loops, and $n_l$ is the number of loops. A noticeable feature of the loop formulation is that it implements on the lattice the continuation to non-integer values of $N$; in particular, for $N\to 0$ it describes the statistics of self-avoiding walks \cite{DeGennes}. The loop model can be solved exactly on the honeycomb lattice \cite{Nienhuis}, on which the loops cannot intersect. The solution yields two critical lines that are defined in the interval $N\in[-2,2]$, coincide at $N=2$, and have critical exponents that were first identified in \cite{DF} as corresponding to the conformal dimensions $\Delta_s$ and $\Delta_\varepsilon$ deduced above for the scattering solutions $P2_\pm$. The two critical lines are often referred to as ``dilute'' and ``dense'' with reference to the loop properties they control\footnote{See \cite{JRS} about the renormalization group properties of the dense phase on lattices allowing for loop intersections.}, and correspond to the solutions $P2_-$ and $P2_+$, respectively. The analogy between loop paths and particle trajectories was originally observed in \cite{Zamo_SAW} for the off-critical case. 

\begin{table}
\centering
\begin{tabular}{c|c|c|c|c|c}
\hline 
Solution & $N$ & $c$& $\Delta_\eta$ & $\Delta_\varepsilon$ & $\Delta_{s}$  \\ 
\hline \hline
$\text{P}1_-$ & $\mathbb{R}$ & $\frac{N}{2}$ & $\frac{1}{2}$ & $\frac{1}{2}$ & $\frac{1}{16}$  \\ 
$\text{P}1_+$ & $\mathbb{R}$ & $N-1$ & $0$ & $1$ & $0$  \\ 
$\text{P}2_{-}$ & $2\cos\frac{\pi}{p}$ & $1-\frac{6}{p (p+1)}$  & $\Delta_{2,1}$ & $\Delta_{1,3}$ & $\Delta_{\frac{1}{2}, 0}$ \\ 
$\text{P}2_{+}$ & $2\cos\frac{\pi}{p+1}$ & $1-\frac{6}{p(p+1)}$ & $\Delta_{1,2}$ & $\Delta_{3, 1}$ & $\Delta_{0, \frac{1}{2}}$  \\
$\text{P}3_{\pm}$ & $2$ & $1$ & $\frac{1}{4b^2}$ & $b^2$ & $\frac{1}{16b^2}$  \\ 
\hline 
\end{tabular} 
\caption{Central charge and conformal dimensions for the solutions of the pure case. The conformal dimensions $\Delta_{\mu,\nu}$ are specified by (\ref{deltamunu}). Other specifications are discussed in the text.
}
\label{table_pure}
\end{table}

\subsection{Critical line at $N=2$ and the BKT phase}
The solutions $P3_\pm$ can be rewritten as
\EQ
\rho_1=\sin\alpha\,,\hspace{1cm}\rho_2=\cos\alpha\,,\hspace{1cm}\phi=-\frac{\pi}{2}\,.
\label{p3alpha}
\EN
The presence of a free parameter ($\alpha$ in the formulation (\ref{p3alpha})) leads to a line of fixed points with $N=2$. This is not surprising since it is well known that the realization of $O(2)$ symmetry with smallest central charge in two-dimensional CFT is provided by the free bosonic theory with action
\EQ
{\cal A}=\frac{1}{4\pi}\int d^2x\,(\nabla\varphi)^2\,,
\label{gauss}
\EN
which indeed describes a line of fixed points with central charge $c=1$ \cite{DfMS}. The energy density field $\varepsilon(x)=\cos 2b\varphi(x)$, with conformal dimension $\Delta_\varepsilon=b^2$, contains the parameter $b$ providing the coordinate along the line of fixed points. It was shown in \cite{paraf} that $\Delta_\eta=1/4b^2$ on this line. Since (\ref{singlet}) gives $S=e^{-i\alpha}$, (\ref{phase}) yields the relation
\EQ
\alpha=\frac{\pi}{2b^2}\,,
\label{phi_b}
\EN
which satisfies $S=-1$ at the point $b^2=1/2$. This is necessary because the theory (\ref{gauss}) is known to possess also a fermionic (Thirring) formulation with action \cite{DfMS}
\EQ
{\cal A}=\int d^2x\,\left[\sum_{i=1,2}(\psi_i\bar{\partial}\psi_i+\bar{\psi}_i\partial\bar{\psi_i})+g(b^2)\psi_1\bar{\psi_1}\psi_2\bar{\psi}_2\right],
\label{Thirring}
\EN
with $b^2=1/2$ corresponding to free fermions ($g(1/2)=0$). The particles $a=1,2$ of the scattering theory correspond to the two neutral fermions in (\ref{Thirring}). $O(2)$ symmetry of the action (\ref{Thirring}) also yields $\Delta_s=1/16b^2$ \cite{paraf}.

\begin{figure}
\begin{center}
\includegraphics[width=9cm]{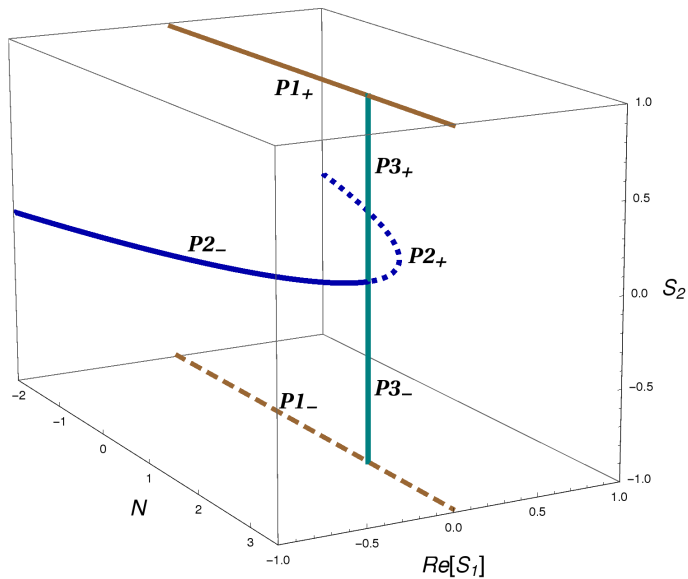}
\caption{Solutions of fixed point equations in the pure case. Those with $S_2=0$ correspond to the critical lines for the dilute (continuous) and dense (dotted) phases of non-intersecting loops. The piece $P3_+$ of the line of fixed points for $N=2$ accounts for the BKT phase of the $XY$ model. The solution $P1_+$ corresponds to the zero temperature critical point of the $N>2$ ferromagnet.
}
\label{pure_space}
\end{center} 
\end{figure}

The intervals $\alpha\in[0,\pi/2]$ and $\alpha\in[\pi/2,\pi]$ correspond to solutions $P3_+$ and $P3_-$, respectively, and have in common the point $\alpha=\pi/2$, which is also the merging point of the solutions $P2_\pm$ (see figure~\ref{pure_space}). Since the field $\varepsilon=\cos 2b\phi$ is irrelevant in the renormalization group sense ($\Delta_\varepsilon=b^2>1$) for $\alpha\in[0,\pi/2]$, $P3_+$ accounts for the BKT phase \cite{BKT} of the $XY$ ferromagnet, i.e. the model with Hamiltonian (\ref{lattice}), $J_{ij}=J>0$ and $N=2$. The point $\alpha=\pi/2$ is the BKT transition point, where the field $\varepsilon$ becomes marginal ($\Delta_\varepsilon=1$). 

Since $\phi$ is fixed in (\ref{p3alpha}), $\rho_1$ does not need to be positive. The part of the $c=1$ line with $b^2<1/2$ (i.e. $\alpha>\pi$) is then also mapped on (\ref{p3alpha}), and corresponds to $P3_+$ or $P3_-$ depending on the sign of $\rho_2$.

\subsection{Free solutions and zero temperature critical point for $N>2$}
The solutions $P1_+$ and $P1_-$ are purely transmissive and correspond to free bosons and fermions, respectively. $P1_-$ can straightforwardly be identified as corresponding to $N$ free neutral fermions, for a total central charge $c=N/2$. For $N=2$ one recovers the $c=1$ free fermion point described by (\ref{gauss}) for $b^2=1/2$, or by (\ref{Thirring}) with $g=0$; this is the contact point between $P1_-$ and $P3_-$ in figure~(\ref{pure_space}). For $N=1$ one recovers the Ising central charge $1/2$. Notice, however, that this Ising point on $P3_-$ does not coincide in scattering space with that on $P2_-$, which is realized non-transmissively: indeed, for $N=1$ there are no particle indices to distinguish between $S_1$, $S_2$ and $S_3$, and only $S=S_1+S_2+S_3=-1$ matters for a free fermion. The value of the conformal dimension $\Delta_s=1/16$ we report in table~\ref{table_pure} for $P1_-$ is that of the multiplet $(s_1,\ldots,s_N)$ containing the spin fields of the $N$ decoupled Ising copies. When comparing with the free fermion point $b^2=1/2$ of $P3_-$ one has to consider that the spin vector field along the $N=2$ line has a specific representation \cite{paraf}, which at $b^2=1/2$ corresponds to the conformal dimension\footnote{See also \cite{AT1,AT2} on spin fields in the fermionic theory (\ref{Thirring}) with $N=2$.} $\Delta_{s_1s_2}=2\Delta_{s_1}=1/8$. 

The solution $P1_+$ can certainly describe $N$ free bosons, namely a theory characterized by the action $\sum_{j=1}^N\int d^2x(\nabla\varphi_j)^2$, with $\Delta_\varepsilon=\Delta_{\varphi^2_j}=0$, and $c=N$. However, the fact that the point $N=2$ of $P1_+$ can also be seen as the limit $b^2\to\infty$ of $P3_+$ (which has $c=1$) says that $P1_+$ must also allow for a different interpretation. We recall that scattering on the line involves position exchange, and mixes statistics with interaction. This is why for $N=2$ the interacting fermions of the theory (\ref{Thirring}) can appear for $b^2\to\infty$ as two free bosons ($S_2=1$). Interaction is known to play a peculiar role also for the critical properties of the $O(N)$ invariant ferromagnet for $N>2$ (see e.g. \cite{Cardy_book}). In this case there is a zero temperature critical point and the scaling properties are described by the non-linear sigma model with action
\EQ
{\cal A}_\textrm{SM}=\frac{1}{T}\sum_{j=1}^N\int d^2x(\nabla\varphi_j)^2\,,\hspace{1cm}\sum_{j=1}^N\varphi_j^2=1\,,
\label{sm}
\EN
where the interaction is introduced by the constraint on the length of the vector $(\varphi_1,\ldots,\varphi_N)$. The theory turns out to be asymptotically free, meaning that the short distance fixed point (which describes the $T=0$ critical point of the ferromagnet) is a theory of free bosons with marginally relevant energy density field ($\Delta_\varepsilon=1$, implying exponentially diverging correlation length as $T\to 0$) and $\Delta_s=\Delta_{\varphi_j}=0$. The constraint gives a central charge $c=N-1$ instead of $N$. The results for $c$ and $\Delta_s$ match those for $N=2$, $b^2\to\infty$. Also $\Delta_\varepsilon=1$ is recovered once we notice that for $b^2>1$ the field $\cos 2b\varphi$ becomes irrelevant, so that the most relevant $O(2)$ invariant field is the marginal one that generates the line of fixed points at $N=2$. 
It is this sigma model interpretation of the solution $P1_+$ that we report in table~\ref{table_pure} together with the other data discussed in this section.

\section{Properties of solutions for the disordered case}
\subsection{Critical lines with varying disorder modulus}
Random fixed points have disorder modulus $\rho_4\neq 0$, and we first consider the solutions $V1_\pm$ of table~\ref{solutions}. They possess the characteristic property that $\rho_4$ spans, as $N$ varies, the range going from 0 (pure case) to the maximal value 1. More precisely, we focus on the solution $V1_-$, which for $N=1$ coincides with the pure Ising model (point $N=1$ of the solution $P2_-$). We know from the renormalization group version (see \cite{Cardy_book}) of the Harris criterion \cite{Harris} that the scaling dimension of weak disorder is twice that of the energy density field of the pure model. Since the pure Ising model has $\Delta_\varepsilon=1/2$, weak disorder is marginal\footnote{It was shown in \cite{DD} to be marginally irrelevant.} at $N=1$. The fact that $\Delta_\varepsilon=\Delta_{1,3}$ along the solution $P2_-$ implies that weak disorder is weakly relevant slightly below $N=1$, and that a random fixed point can be found perturbatively in this region; such a perturbative analysis is similar to that for $q\to 2^+$ in the Potts model \cite{Ludwig,DPP} and was performed in \cite{Shimada}. More generally, we conclude that the branch of the solution $V1_-$ extending from $N=1$, where $\rho_4=0$, down to
\EQ
N_*=\sqrt{2}-1=0.414..\,,
\label{nstar}
\EN
where $\rho_4=1$, is a line of stable (infrared) fixed points. The existence of a lower endpoint $N_*$ for the infrared critical line was argued in \cite{Shimada}, where the estimate $N_*\approx 0.26$ was obtained in the two-loop approximation. The subsequent estimate $N_*\approx 0.5$ obtained in \cite{SJK} within a numerical transfer matrix study is not far from our exact result.

For $N>1$ the solution $V1_-$ becomes a line of unstable fixed points, weak disorder becoming irrelevant. Indeed the pure model has $\Delta_\varepsilon>1/2$ both for $1<N<2$, where it corresponds to $P2_-$ with $\Delta_\varepsilon=\Delta_{1,3}$, and for $N>2$, where it corresponds to $P1_+$ with $\Delta_\varepsilon=1$. 

\begin{figure}
\begin{center}
\includegraphics[width=9cm]{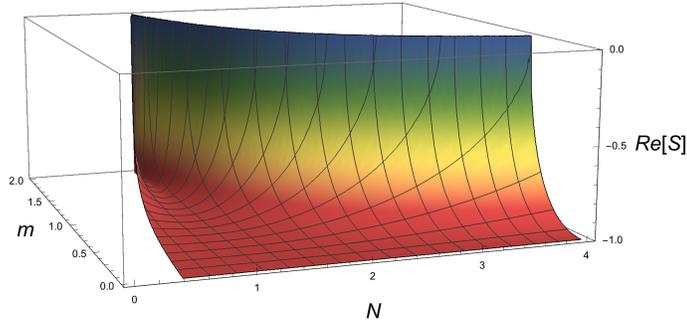}
\caption{The scattering phase (\ref{singlet}) exhibits superuniversality ($N$-independence) in the limit $m\to 0$ corresponding to the solution $V1_-$.  
}
\label{super}
\end{center} 
\end{figure}

Figure~\ref{super} shows the real part of the scattering phase (\ref{singlet}) for the solution (\ref{V1m}) of the fixed point equations, which in the limit $m\to 0$ becomes $V1_-$. We see (and it can be checked analytically) that $S$ becomes $N$-independent {\it at} $m=0$, meaning that the symmetry sector of the superposition $\sum_{a,i}a_ia_i$, which is invariant under $O(N)$ transformations and replica permutations, becomes superuniversal along the line of fixed points $V1_-$. Since the energy density field $\varepsilon(x)$ belongs to this symmetry sector, the dimension $\Delta_\varepsilon$ is expected to keep its $N=1$ (pure Ising) value $1/2$ along the line $V1_-$. This is analogous to the result \cite{random,DT2} recalled in the introduction for the correlation length critical exponent $\nu=[2(1-\Delta_\varepsilon)]^{-1}$ in the random bond $q$-state Potts ferromagnet. On the other hand, the spin field does not fall into the superuniversal sector and its dimension $\Delta_s$ is expected to vary along the critical line $V1_-$. This dimension was measured in \cite{SJK} for $N=0.55$ on the infrared fixed line and found to be consistent with the two-loop result of \cite{Shimada}. 

\subsection{Critical lines with maximal disorder modulus}
\begin{figure}
\begin{center}
\includegraphics[width=8cm]{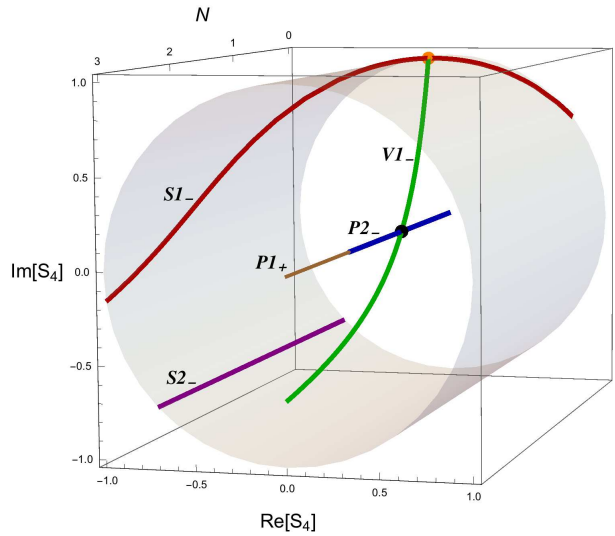}
\caption{Critical lines in the space of the two disorder parameters. The pure case corresponds to $S_4=0$, while lines that are strongly disordered for any $N$ lie on the cylinder $|S_4|=\rho_4=1$.  
}
\label{total}
\end{center} 
\end{figure}

\begin{figure}
\begin{center}
\includegraphics[width=11cm]{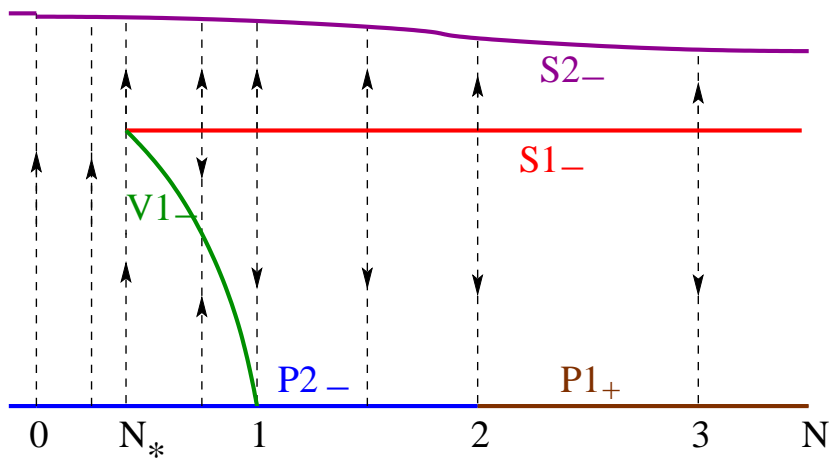}
\caption{Schematic illustration of lines of fixed points and renormalization group flows between them. The critical pure model is at the bottom ($P2_-$, $P1_+$), and is connected to the Nishimori-like multicritical line $S1_-$ by the line of stable fixed points $V1_-$, which spans the interval $N\in(N_*=\sqrt{2}-1,1)$. $S2_-$ is a line of zero temperature stable fixed points. For $N=1$ the renormalization group trajectories correspond to the phase boundary of figure~\ref{phd}.   
}
\label{flows}
\end{center} 
\end{figure}

The last class of solutions in table~\ref{solutions} is characterized by a maximal value $\rho_4=1$ of the disorder modulus and corresponds to critical lines that stay strongly disordered as $N$ varies. We focus on the solutions $S1$ and $S2$, whose range of definition includes $N$ positive. Continuing the physical considerations of the previous subsection, we see that the infrared branch of the solution $V1_-$ coincides at its endpoint $N_*$ with the solution $S1_-$ (see figure~\ref{total}). This indicates that for $N>N_*$ $S1_-$ is a line of unstable fixed points. There are renormalization group trajectories that emanate from this line and end on the line $V1_-$ for $N\in(N_*,1)$, and on the pure model for $N>1$ (figure~\ref{flows}). The point $N=1$ on the line $S1_-$ then corresponds to the Nishimori multicritical point $M$ in the Ising phase diagram of figure \ref{phd}. A line of strongly disordered fixed points extending for $N>N_*$ was indeed observed numerically in \cite{SJK}, and its universal properties at $N=1$ were found to be quantitatively consistent with those of the Nishimori point. We refer to the portion $N>N_*$ of the critical line $S1_-$ as a line of Nishimori-like multicritical points, implying by this that, from the renormalization group point of view, the points on this line play a role analogous to that of the Nishimori point at $N=1$, while the lattice gauge symmetry characteristic of the $\pm J$ Ising model \cite{Nishimori} is in general absent for $N\neq 1$. We also notice that the symmetry sector corresponding to the scattering phases (\ref{Sigma}) is $N$-independent along $S1_-$. 

When we say that for $N>1$ there is a renormalization group flow from the Nishimori-like critical line $S1_-$ to the pure model, we mean solution $P2_-$ for $N\in(1,2)$, and solution $P1_+$ for $N>2$. At $N=2$ the flow from the Nishimori-like point can end on any point of the line of fixed points $P3_+$. This explains the BKT phase observed in numerical studies (see e.g. \cite{APV,OYSE}) of the disordered $XY$ model\footnote{The model studied in \cite{APV,OYSE} is the random phase XY model, which has ${\bf s}_i=(\cos\alpha_i,\sin\alpha_i)$, nearest neighbor interaction $-\cos(\alpha_i-\alpha_j+A_{ij})$, and random variables $A_{ij}$ drawn from a distribution $P(A_{ij})\propto e^{-A^2_{ij}/\sigma}$; $\sigma$ replaces $1-p$ in figure~\ref{phd}. Since our formalism relies only on symmetry, it applies also to this type of disorder.}. These studies find a phase diagram analogous to that of figure~\ref{phd}, with a BKT phase replacing the ferromagnetic phase.

It is a consequence of our analysis that, having $\rho_4=1$, the Nishimori-like critical line never approaches the pure case. It follows that the zero temperature fixed point of figure~\ref{phd} must have the same property, and must also belong to the class of solutions with $\rho_4=1$. Excluding $S3$ and $S4$, which are not defined for $N$ positive, the main candidate for a zero temperature critical line appears to be $S2_-$, which differs from $S1_-$ only for the $\theta$-dependence. Notice that this solution actually is completely $N$-independent, so that conformal dimensions and critical exponents should not vary along it. Hence, $S2_-$ is expected to be a line of infrared fixed points for any $N$. It follows that for $-2< N<N_*$, where the pure solution is unstable and the solution $V1_-$ is not defined, the flow from the pure model should end directly on  the zero temperature line $S2_-$. 

The pattern of disorder driven flows discussed in this section is schematically summarized in figure~\ref{flows}. For $N_*<N<1$ there is a sequence of three flows between four fixed points ($P2_-\rightarrow V1_-\leftarrow S1_-\rightarrow S2_-$) that accounts for the pattern observed numerically at $N=0.6$ in \cite{SJK}. The flows observed in \cite{SJK} in the ranges $0<N<N_*$ and $1<N<2$ also match those following from our identifications. Our results also explain why the flow from the pure model to the Nishimori-like point is not observed in \cite{SJK} for $N>2$. The reason is that for $N>2$ the hexagonal lattice loop model studied in that paper cannot see the critical point of the pure $O(N)$ model\footnote{It sees instead a critical point with $Z_3$ symmetry inherited from the lattice structure \cite{GBW}.}. Indeed, we saw that this corresponds to $P1_+$, a purely transmissive ($S1=S3=0$) solution that is not realized on the hexagonal lattice. 

The fact that for $N>2$ both the pure fixed point and that corresponding to $S2_-$ fall at zero temperature suggests that in this range also the Nishimori-like fixed point should be at zero temperature. In this respect it is interesting to notice that {\it two} zero temperature disordered fixed points were found in \cite{SJK} for $N=8$. 

One implication of our results is that the large distance physics of self-avoiding walks in a random medium ($N=0$) is controlled by the fixed point on $S2_-$. Consistently, the amplitudes $S_2$ and $S_5$ corresponding to intersecting walks vanish for this solution.

\section{Conclusion}
We obtained all solutions of the exact fixed point equations yielded by the scale invariant scattering formalism for replicated $O(N)$ symmetry, which we then specialized to the limit of zero replicas that implements quenched disorder. The method is particularly powerful since, as $N$ varies, the different lines of fixed points are obtained exactly and can be located within the same parameter space. This is relevant already for the critical lines of the pure case, which fall into a subspace spanned by $N$ and two interaction parameters. We have shown, in particular, how the lines of fixed points ruling the critical behaviors of the dilute and dense phases of non-intersecting loops ($-2\leq N\leq 2$) are connected to the zero temperature fixed points for $N>2$ via the BKT line at $N=2$. Our solutions also show that for $N>2$ the pure case corresponds to purely intersecting paths, and clarify why such fixed points are not observed in studies of the loop gas on the hexagonal lattice. 

Disorder introduces two additional parameters that we call $\rho_4$ and $\theta$. The ``disorder modulus'' $\rho_4$ vanishes in the pure limit, while its maximal value 1 selects a subspace of solutions containing the  stable zero temperature critical line and a line of Nishimori-like multicritical points (coinciding with the Nishimori point at $N=1$). We identified the main candidate for the stable zero temperature critical line with a completely $N$-independent solution, and it would be very interesting to test this conjectured superuniversality of critical exponents by numerical simulations in the disordered model for different values of $N$. 

A solution with $\rho_4$ varying with $N$ describes a line of infrared fixed points going from $N_*=\sqrt{2}-1=0.414..$ on the Nishimori-like line to the pure Ising fixed point at $N=1$. Along this infrared line we find superuniversality in the symmetry sector containing the correlation length critical exponent $\nu$, a prediction that could be tested through the numerical transfer matrix approach of \cite{SJK}. The estimate $N_*\approx 0.5$ obtained in that paper is consistent with our exact result. 

For $-2<N<N_*$ we find that weak disorder drives the system directly to the zero temperature fixed point, consistently with the pattern observed numerically in \cite{SJK}. In particular this occurs for the case $N=0$ corresponding to self-avoiding walks in a disordered medium. Hence, if complete superuniversality of the zero temperature critical line is confirmed, it should also yield the exponents of the disordered polymer. For $N\geq 1$ weak disorder is irrelevant and there is renormalization group flow from the Nishimori-like line to the pure model. 

In general, the scale invariant scattering approach to random criticality in two dimensions initiated in \cite{random} has shown, when such an outcome no longer seemed likely, that exact results can be obtained for this problem. In particular, it shows that random fixed points possess conformal invariance, which in the particle description is responsible for the elasticity of scattering processes. It also emerges that the CFT's of random fixed points allow for superuniversal (symmetry independent) sectors, a circumstance that has no counterpart in the pure case. Probably this peculiarity has contributed to the difficulty of identifying CFT's of random criticality but, having been recognized, could also serve as a guide towards the remaining goal -- completing for the random solutions a table of conformal data similar to table~2 above for the pure ones.



\begin{thebibliography}{99}

\bibitem{Cardy_book} J. Cardy, Scaling and renormalization in statistical physics, Cambridge, 1996.
\bibitem{BPZ} A.A. Belavin, A.M. Polyakov and A.B. Zamolodchikov, Nucl. Phys. B 241 (1984) 333.
\bibitem{DfMS} P. Di Francesco, P. Mathieu and D. Senechal, Conformal field theory, Springer-Verlag, New York, 1997.
\bibitem{paraf} G. Delfino, Annals of Physics 333 (2013) 1.
\bibitem{fpu} G. Delfino, Annals of Physics 360 (2015) 477.
\bibitem{DT1} G. Delfino and E. Tartaglia, Phys. Rev. E 96 (2017) 042137.
\bibitem{GRZ} V. Gorbenko, S. Rychkov and B. Zan, SciPost Phys. 5, 050 (2018).
\bibitem{DHJSS} Y. Deng, Y. Huang, J.L. Jacobsen, J. Salas, and A.D. Sokal, Phys. Rev. Lett. 107 (2011) 150601.
\bibitem{Huang} Y. Huang, K. Chen, Y. Deng, J.L. Jacobsen, R. Koteck\'y, J. Salas, A.D. Sokal and J.M. Swart, Phys. Rev. E 87 (2013) 012136.
\bibitem{LDJS} J.-P. Lv, Y. Deng, J.L. Jacobsen and J. Salas, J. Phys. A: Math. Theor. 51 (2018) 365001.
\bibitem{random} G.~Delfino, Phys. Rev. Lett. 118 (2017) 250601.
\bibitem{AW} A. Aizenman and J. Wehr, Phys. Rev. Lett. 62 (1989) 2503.
\bibitem{HB} K. Hui and A.N. Berker, Phys. Rev. Lett. 62 (1989) 2507. 
\bibitem{CFL} S. Chen, A.M. Ferrenberg and D.P. Landau, Phys. Rev. Lett. 69 (1992) 1213; Phys. Rev. E 52 (1995) 1377. 
\bibitem{DW} E. Domany and S. Wiseman, Phys. Rev. E 51 (1995) 3074.
\bibitem{KSSD} M. Kardar, A.L. Stella, G. Sartoni and B. Derrida, Phys. Rev. E 52 (1995) R1269.
\bibitem{CJ} J. Cardy and J.L. Jacobsen, Phys. Rev. Lett. 79 (1997) 4063.
\bibitem{CB2} C. Chatelain and B. Berche, Phys. Rev. E 60 (1999) 3853.
\bibitem{OY} T. Olson and A. P. Young, Phys. Rev. B 60 (1999) 3428.
\bibitem{JP} J.L. Jacobsen and M. Picco, Phys. Rev. E 61 (2000) R13.
\bibitem{Jacobsen_multiscaling} J.L. Jacobsen, Phys. Rev. E 61 (2000) R6060(R).
\bibitem{AdAI} J.-Ch. Angl\`es d'Auriac and F. Igloi, Phys. Rev. Lett. 90 (2003) 190601.
\bibitem{DL1} G. Delfino and N. Lamsen, JHEP 04 (2018) 077.
\bibitem{Shimada} H. Shimada, Nucl. Phys. B 820 (2009) 707.
\bibitem{SJK} H. Shimada, J.L. Jacobsen and Y. Kamiya, J. Phys. A 47 (2014) 122001.
\bibitem{PHP} M. Picco, A. Honecker and P. Pujol, J. Stat. Mech. (2006) P09006. 
\bibitem{HPtPV} M. Hasenbusch, F. Parisen Toldin, A. Pelissetto and E. Vicari, Phys. Rev. E 77 (2008) 051115.
\bibitem{Nishimori} H. Nishimori, Prog. Theor. Phys. 66 (1981) 1169.
\bibitem{ELOP} R.J. Eden, P.V. Landshoff, D.I. Olive and J.C. Polkinghorne, The analytic S-matrix, Cambridge, 1966.
\bibitem{D_09} G. Delfino, Nucl. Phys. B 807 (2009) 455.
\bibitem{Zamo_cth} A.B. Zamolodchikov, Sov. J. Nucl. Phys. 46 (1987), 1090.
\bibitem{DeGennes} P.G. De Gennes, Phys. Lett. A 38 (1972) 339.
\bibitem{Nienhuis} B. Nienhuis, J. Stat. Phys. 34 (1984) 731.
\bibitem{DF} Vl.S. Dotsenko and V.A. Fateev, Nucl. Phys. B 240 (1984) 312.
\bibitem{JRS} J.L. Jacobsen, N. Read and H. Saleur, Phys. Rev. Lett. 90 (2003) 090601.
\bibitem{Zamo_SAW} A.B. Zamolodchikov, Mod. Phys. Lett. A 6 (1991) 1807.
\bibitem{BKT} V.L. Berezinskii, Sov. Phys. JETP, 32 (1971) 493.

J.M. Kosterlitz and D.J. Thouless, J. Phys. C 6 (1973) 1181.
\bibitem{AT1} G. Delfino, Phys. Lett. B 450 (1999) 196.
\bibitem{AT2} G. Delfino and P. Grinza, Nucl. Phys. B 682 (2004) 521.
\bibitem{Harris} A.B. Harris, J. Phys. C 7 (1974) 1671.
\bibitem{DD} V.S. Dotsenko and Vl. S. Dotsenko, Sov. Phys.--JETP Lett. 33 (1981) 37; Adv. Phys. 32 (1983) 129.
\bibitem{Ludwig} A.W.W. Ludwig, Nucl. Phys. B 330 (1990) 639.
\bibitem{DPP} V. Dotsenko, M. Picco and P. Pujol, Nucl. Phys. B 455 (1995) 701.
\bibitem{DT2} G. Delfino and E. Tartaglia, J. Stat. Mech. (2017) 123303.
\bibitem{APV} V. Alba, A. Pelissetto and E. Vicari, J. Stat. Mech. (2010) P03006.
\bibitem{OYSE} Y. Ozeki, S. Yotsuyanagi, T. Sakai and Y. Echinaka, Phys. Rev. E 89 (2014) 022122.
\bibitem{GBW} W. Guo, H.W.J. Blote and F.Y. Wu, Phys. Rev. Lett. 85 (2000) 3874.



\end{thebibliography}
\end{document}